\documentstyle[aps,prl,graphicx]{revtex}

\input psfig.sty

\newcommand{\statetwo}{\protect{$\left | 2 \right>$}}
\newcommand{\stateones}{\protect{$\left | 1 \right>\ $}}
\newcommand{\statetwos}{\protect{$\left | 2 \right>\ $}}

\begin{document}
\twocolumn
\wideabs{
\title{Vortex precession in Bose-Einstein condensates: observations with filled and empty cores}
\author{B.~P. Anderson\cite{qpdNIST}, P.~C. Haljan, C.~E.
 Wieman, and E.~A. Cornell\cite{qpdNIST}}
\address{JILA, National Institute of Standards and Technology and Department of Physics, \\
University of Colorado, Boulder, Colorado 80309-0440}
\date{May 22, 2000}

\maketitle

\begin{abstract}

We have observed and characterized the dynamics of singly
quantized vortices in dilute-gas Bose-Einstein condensates. Our
condensates are produced in a superposition of two internal states
of $^{87}$Rb, with one state supporting a vortex and the other
filling the vortex core. Subsequently, the state filling the core
can be partially or completely removed, reducing the radius of the
core by as much as a factor of 13, all the way down to its bare
value. The corresponding superfluid rotation rates, evaluated at
the core radius, vary by a factor of 150, but the precession
frequency of the vortex core about the condensate axis changes by
only a factor of two.\\

PACS number(s): 03.75.Fi, 67.90.+z, 67.57.Fg, 32.80.Pj
\end{abstract}


}

The dynamics of quantized vortices in superfluid helium and
superconductors have been fascinating and important research areas
in low-temperature physics \cite{Tilley1990aDonnelly1991a}.
Continued study of vortex dynamics may, for example, lead to a
better understanding of energy dissipation in these systems
\cite{Thouless1999a}. Work on optical vortices has also become an
active area of research \cite{Swartzlander1992aKivshar1998a}. More
recently, demonstrations of the creation of quantized vortices in
dilute-gas Bose-Einstein condensates (BEC)
\cite{Matthews1999a,Madison2000a} have proven to be striking
examples of the similarities between the condensed matter,
optical, and dilute-gas quantum systems. Because of the
observational capabilities of dilute-gas BEC experiments and the
ability to manipulate the quantum wavefunction of the condensates,
these systems provide a unique approach to the study of quantized
vortices and their dynamics.  This paper reports direct
observations and measurements of singly quantized vortex core
precession in a BEC.

Numerous theoretical papers have explored the expected stability
and behavior of vortices in BEC
\cite{Rokhsar1997a,Butts1999aPu1999aGarcia-Ripoll1999a,Bolda1998x,Skryabin2000a,Jackson1999b,Svidzinsky1998b,Fedichev1999a,Perez-Garcia1999aGarcia-Ripoll1999b}.
One interesting predicted effect is vortex core precession about
the condensate axis
\cite{Rokhsar1997a,Bolda1998x,Skryabin2000a,Jackson1999b,Svidzinsky1998b,Fedichev1999a}.
Radial motion of the core within the condensate can also occur,
and may be understood as being due to energy dissipation and
damping processes.

Core precession may be described in terms of a Magnus effect -- a
familiar concept in fluid dynamics and superfluidity
\cite{Tilley1990aDonnelly1991a}.  An applied force on a rotating
cylinder in a fluid leads to cylinder drift (due to pressure
imbalances at the cylinder surface) that is orthogonal to the
force. Analogously, a net force on a vortex core in a superfluid
results in core motion perpendicular to both the vortex
quantization axis and the force.  In the condensate vortex case,
these forces can be due to density gradients within the
condensate, for example, or the drag due to thermal atoms. The
density-gradient force may be thought of as one component of an
effective buoyancy: just as a bubble in a fluid feels a force
anti-parallel to the local pressure gradient, a vortex core in a
condensate will feel a force towards lower condensate densities.
The total effective buoyancy, however, is due less to displaced
mass (the ``bubble'') than it is to dynamical effects of the
velocity-field asymmetry, which in turn is a consequence of a
radially offset core. Typically, the total buoyancy force is
towards the condensate surface, and the net effect is a precession
of the core around the condensate axis via the Magnus effect. Drag
due to the motionless (on average) thermal atoms opposes core
precession, causing the core to spiral outwards towards the
condensate surface. In the absence of this drag (for temperature
$\sim0$), radial drift of the core may be negligible.

Our techniques for creating and imaging a vortex in a coupled
two-component condensate are described in
refs.~\cite{Matthews1999a,Williams1999c}. The two components are
the \protect{$\left | F=1,m_{F}=-1 \right>$} and \protect{$\left |
F=2,m_{F}=1 \right>$} internal states of the $^{87}$Rb atom,
henceforth labeled as states \stateones and \statetwos
respectively.  We start with a condensate of $10^6$ \statetwos
atoms, confined in a spherical potential with oscillator frequency
7.8 Hz.  A near-resonant microwave field causes some of the
\statetwos atoms to convert to \stateones atoms. The presence of a
rotating, off-resonant laser beam spatially modulates the
amplitude and phase of the conversion. The net result is a
conversion of about half of the sample into an annular ring of
\stateones atoms with a continuous quantum phase winding from 0 to
2$\pi$ about the circumference -- a singly quantized vortex. The
balance of the sample remains in the non-rotating \statetwos state
and fills the vortex core. With resonant light pressure we can
selectively remove as much of the core material as we desire. In
the limit of complete removal, we are left with a
single-component, bare vortex state.

In this bare-core limit, the core radius is on the order of the
condensate healing length $\xi=(8\pi n_{0}a)^{-1/2}$, where
$n_{0}$ is the peak condensate density and $a$ is the scattering
length.  For our conditions, $\xi=0.65~\mu$m, well under our
imaging resolution limit. The bare core can be observed after
ballistic expansion \cite{Madison2000a} of the condensate, but
this is a destructive measurement.  On the other hand, if we leave
some of the \statetwo-state atoms filling the core, the pressure
of the filling material opens up the radius of the \stateones
vortex core to the point where we can resolve the core in a time
series of nondestructive phase-contrast images.

\begin{figure}[p]
\begin{center}
\psfig{figure=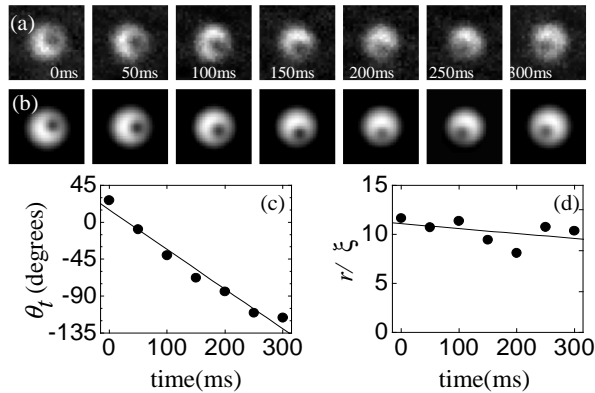,width=1\linewidth,clip=}
\end{center}
\caption {(a) Seven successive images of a condensate with a
vortex and (b) their corresponding fits.  The 75-$\mu$m-square
nondestructive images were taken at the times listed, referenced
to the first image. The vortex core is visible as the dark region
within the bright condensate image. (c) The azimuthal angle of the
core is determined for each image, and is plotted vs.\ time held
in the trap.  A linear fit to the data indicates a precession
frequency of 1.3(1) Hz for this data set. (d) Core radius $r$ in
units of healing length $\xi$. The line shown is a linear fit to
the data.}\label{filledimage}
\end{figure}

\emph{Filled-core dynamics --} We first discuss vortex dynamics in
two-component condensates, where $10\% - 50\%$ of the atoms were
in the \statetwos fluid filling the \stateones vortex core.  We
took successive images of the \stateones atoms in the magnetic
trap, with up to 10 images of each vortex.  The vortex core is
visible as a dark spot in a bright \stateones distribution, as
shown in Fig.~\ref{filledimage}(a). Instabilities in our vortex
creation process usually resulted in the creation of off-center
vortex cores, allowing us to observe precession of the cores. We
observed precession out to $\sim$ 2 s, after which the \statetwos
fluid had decayed to the point that the vortex core was too small
to be observed in the trapped condensate.

The recorded profile of each trapped condensate was fit with a
smooth Thomas-Fermi distribution. Each vortex core profile was fit
with a Gaussian distribution to determine its radius and position
within the condensate. From the fits, we determined the overall
radius $R_{t}$ of the trapped condensate (typically 22 $\mu$m),
the HWHM radius $r$ of the filled vortex core, and the
displacement $d_{t}$ and angle $\theta_{t}$ of the core center
with respect to the condensate center.  Core angles and radii for
the images in Fig.~\ref{filledimage}(a) are shown in
Figs.~\ref{filledimage}(c) and (d). The vortex core is seen
precessing in a clockwise direction, which is the same direction
as the vortex fluid flow around the core.
\noindent The angular precession frequency was determined from the
time dependence of $\theta_{t}$ (Fig.~\ref{filledimage}(c)). This
and other similar data sets showed no reproducible radial motion
of the core over the times and parameters examined. However,
consistent decrease in the size of the core was observed, which we
interpret as being due to known decay of the \statetwos fluid
through inelastic atomic collisional processes.

\begin{figure}[p]
\begin{center}
\psfig{figure=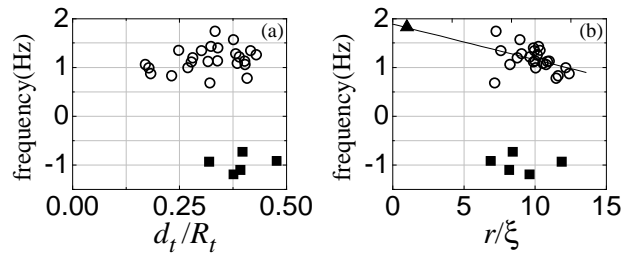,width=1\linewidth,clip=}
\end{center}
\caption{Compiled data for filled vortex core precession, with
each data point extracted from a series (as in
Fig.~\ref{filledimage}) of nondestructive images of a single
vortex.  Precession frequency is plotted vs.\ (a) core
displacement $d_{t}$ in units of condensate radius $R_{t}$, and
(b) core radius $r$ in units of healing length $\xi$. Circles
correspond to positive frequencies and filled squares to negative
frequencies.  (Positive frequency is defined as core precession
having the same handedness as the vortex angular momentum.)  The
triangle at $r=\xi$ shows for reference the average measured
precession frequency of many bare vortices (see text and
Fig.~\ref{baredynamics}(b)). A line is drawn as a guide to the
trend in frequency vs.\ core size. }\label{filledcompiled}
\end{figure}

For each data set, we determined a mean core radius and
displacement. The data cover a range of core radii ($r=7\xi$ to
$13\xi$), displacements ($d_{t} = 0.17R_{t}$ to $0.48R_{t}$), and
percentage of atoms in the core ($10\%$ to $50\%$).  Except for a
few ``rogue vortices'' (discussed below), the measured precession
frequencies are clustered around 1.4 Hz, as shown in
Fig.~\ref{filledcompiled}, precessing in the same direction as the
vortex fluid rotation. The data (Fig.~\ref{filledcompiled}(a))
suggest a slight increase in frequency for cores further from the
condensate center, in qualitative agreement with two-dimensional
numerical simulations for two-component condensates
\cite{HollandPrep}. We also see (Fig.~\ref{filledcompiled}(b)) a
slight decrease in precession frequency for larger cores.

As indicated in Fig.~\ref{filledcompiled}, a few vortex cores
exhibited precession opposite to that of the fluid flow, but with
absolute frequencies similar to the bulk of the data.  The quality
of the corresponding vortex images was routinely lower than for
the positive-frequency precession points, with vortices looking
more like crescents and ``D'' shaped objects rather than like the
images of Fig.~\ref{filledimage}.  We speculate that this
``inverse precession'' may be due to distorted configurations of
vortices in the two-component condensate in which vortex dynamics
are not as well understood.  Such configurations have received
recent theoretical attention
\cite{Perez-Garcia1999aGarcia-Ripoll1999b}.

\begin{figure}[p]
\begin{center}
\psfig{figure=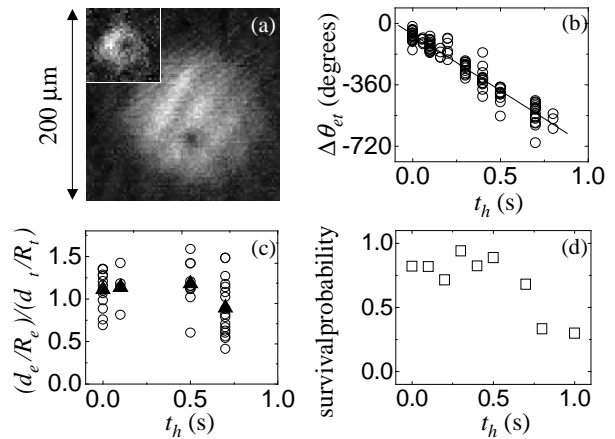,width=1\linewidth,clip=}
\end{center} \caption {(a) Ballistic expansion image of a vortex
after all \statetwos atoms have been removed. The dark spot is the
bare vortex core. Inset -- the corresponding, preceding in-trap
nondestructive image of the partially-filled core. (b) Angular
differences $\Delta\theta_{et}$ between vortex cores from the
in-trap and expansion images, plotted against hold time $t_{h}$ in
the magnetic trap.  The line is a fit through the data, indicating
a bare core precession frequency of 1.8(1) Hz. (c) Radial core
motion is determined by $(d_{e}/R_{e})/(d_{t}/R_{t})$, the ratio
of the fractional core displacements from the expansion and
in-trap images of each data set. The data are shown as open
circles, with the average of all data at each given hold time
plotted as a filled triangle. (d) Survival probability of a
vortex, defined as the conditional probability for observing a
vortex in an expanded image given the observation of a vortex in
the corresponding, pre-expansion in-trap image. Bare cores are not
observed for hold times $t_{h} > 1.1$~s.}\label{baredynamics}
\end{figure}

\newpage \emph{Bare core dynamics --}  To examine the dynamics of bare
vortices, our procedure consisted of taking a nondestructive
phase-contrast picture of the partially filled \stateones vortex
distribution (Fig.~\ref{baredynamics}(a) inset), as previously
discussed, followed by complete removal of the core filling
\cite{correlations}. We then held the bare vortex in the trap for
a variable hold time $t_{h}$, after which the condensate was
released from the trap. We took a final near-resonance
phase-contrast image \cite{imagetechnique} of the atomic
distribution (Fig.~\ref{baredynamics}(a)) after the condensate had
ballistically expanded by a factor of $\sim 3.5$
\cite{expandtechnique} and the core had expanded
\cite{Dalfovo2000a} to a fit radius of $\sim9 \mu$m.

Displacements $d_{t}$ and angular positions $\theta_{t}$ of the
cores for the in-trap images were extracted as described before.
The images of the expanded clouds were fit with identical
distributions, and the Thomas-Fermi radius $R_{e}$ of the expanded
cloud and the vortex core displacement $d_{e}$ and angle
$\theta_{e}$ were obtained for each image. For each pair of
images, we determined the angular difference $\Delta\theta_{et}
\equiv \theta_{e} - \theta_{t}$ between the cores in the expanded
and in-trap images.  We also determined the core displacement
ratio $d_{e}/d_{t}$, an indicator of the radial motion of the core
during the hold time $t_{h}$.

From the measurements of $\Delta\theta_{et}$ at different hold
times $t_{h}$ (Fig.~\ref{baredynamics}(b)), we find a bare core
precession frequency of 1.8(1) Hz, which is slightly faster than
the precession of filled cores and consistent with the trend shown
in Fig.~\ref{filledcompiled}(b) for filled cores.  To emphasize
that our measurements of filled and empty cores are different
limits in a continuum of filling material, we indicate the
measured bare core precession frequency in
Fig.~\ref{filledcompiled}(b) as a point at $r = \xi$.

From Fig.~\ref{filledcompiled}(b) it is apparent that the
structure and content of the vortex core have a relatively modest
effect on precession frequency. One can calculate, for instance,
the fluid rotation rate $\nu_{r}$ at the inner core radius. The
value of $\nu_{r}$ is given by the quantized azimuthal superfluid
velocity evaluated at the radius of the core, divided by the
circumferential length at that radius. For the bare-core vortex
(the triangle in Fig.~\ref{filledcompiled}(b)) $\nu_{r}$ is about
260 Hz, while for the largest filled cores of
Fig.~\ref{filledcompiled}(b) (for which nearly half of the sample
mass is composed of core filling), $\nu_{r}$ is only about 1.7 Hz.
Thus between vortices whose inner-radius fluid rotation rates vary
by a factor of 150, we see only a factor of 2 difference in
precession frequency.

The slower precession of filled cores can be understood in terms
of our buoyancy picture.  Due to its slightly smaller scattering
length, \statetwos fluid has negative buoyancy with respect to
\stateones fluid, and consequently tends to sink inward towards
the center of the condensate \cite{sinktheoryexperiment}.  With
increasing amounts of \statetwos material in the core, the inward
force on the core begins to counteract the outward buoyancy of the
vortex velocity field, resulting in a reduced precession velocity.
It is predicted that with a filling material of sufficiently
negative buoyancy in the core, the core precession may stop or
even precess in a direction
\noindent  opposite to the direction of the fluid flow
\cite{HollandPrep}, but our data do not reach this regime.

Various theoretical techniques involving two- and
three-dimensional numerical and analytical analyses have been
explored to calculate the precession frequency of a vortex core
within a condensate
\cite{Jackson1999b,Svidzinsky1998b,Fedichev1999a,Lundh1999a}. We
briefly compare those most readily applied to our physical
parameters, assuming a spherical, single-component condensate with
$3\times10^{5}$ atoms in a non-rotating trap. Where relevant, we
assume a core displacement of $d_{t} = 0.35R_{t}$ which is near
the center of our measurement range.  A two-dimensional
hydrodynamic image charge analysis has been analytically explored
in the non-interacting \cite{Fedichev1999a} and weakly-interacting
(Thomas-Fermi) \cite{Lundh1999a} limits. The latter of these
predicts a bare core precession frequency of $\sim 0.8$ Hz.
Svidzinsky and Fetter's  two-dimensional \cite{Svidzinsky1998b}
and three-dimensional \cite{Svidzinskyposter} solution to the
Gross-Pitaevskii equation predicts a precession frequency of $\sim
1.3$ Hz.  Jackson \emph{et.~al.} \cite{Jackson1999b} have obtained
results in close agreement with this analytical solution using a
numerical solution to the Gross-Pitaevskii equation. Finally, a
two-dimensional simulation by McGee and Holland \cite{HollandPrep}
using a steepest-descents technique predicts a precession
frequency of 1.2 Hz.

Measurement of $d_{e}/d_{t}$ for different hold times $t_{h}$
shows the radial motion of the bare cores and is a probe of energy
dissipation of the vortex states.  The plot of
Fig.~\ref{baredynamics}(c) displays no trend of the core towards
the condensate surface during $t_{h}$, indicating that thermal
damping is negligible on the 1 s timescale \cite{T0}. However, a
peculiar feature of our bare core measurements is the lack of
visible cores in the expanded pictures for hold times of
$t_{h}\gtrsim$ 1 s (Fig.~\ref{baredynamics}(d)). The absence of
observed radial motion of the core suggests that the vanishing of
the vortex is not due to thermal damping (which would result in
the core moving towards and annihilating at the condensate
surface). Possible explanations are that the vortex core may tilt
or bend along its length \cite{Svidzinskyposter}, suppressing
contrast in optical depth below our signal-to-noise threshold.

The hypothesis that the vortices are disappearing due to imaging
limitations rather than true decay is supported by an alternative
assay for the presence of vortices:  It has been predicted that
the presence of a vortex will break the degeneracy of the $m=\pm2$
condensate surface modes \cite{modes}. We have experimentally
verified this prediction \cite{HaljanQELS} and used this technique
to verify that the vortex continues to be present in the
condensate well after we can no longer image a core in expansion.
These measurements will be the topic of a future paper
\cite{HaljanPrep}.

Through a combination of destructive and nondestructive imaging
techniques we have obtained measurements of vortex dynamics in
bare- and filled-core vortices in dilute-gas BEC. Vortex
precession frequencies show only modest dependence on the radius
and content of the vortex core. We plan to extend these studies to
investigate higher-order dynamical behavior and to characterize
the dissipative effects of finite temperatures.

We gladly acknowledge helpful discussions with Murray Holland and
Sarah McGee.  This work was supported by funding from NSF, ONR,
and NIST.


\begin{thebibliography}{10}
\bibitem{correlations} The core-filling material is removed slowly
enough (100 ms) that the vortex core has time to shrink
adiabatically to its final bare size, but rapidly enough that the
initial location of the bare core is well correlated with the
location of the original filled core.
\bibitem{imagetechnique} We use phase-contrast imaging for both the in-trap and
expansion images to simplify imaging procedures.  When imaging the
expanded atom cloud, we obtain good signal when the probe is
detuned 3 linewidths from resonance.
\bibitem{expandtechnique} Because of our low trapping frequencies
(7.8 Hz), expansions are correspondingly slow.  We get larger
final spatial distributions by giving the atoms a 6-ms preliminary
``squeeze''at high spring constant, followed by a 50-ms expansion
period.
\bibitem{T0} Our measurements are performed at temperatures $T/T_{c} < 0.6$,
where $T_{c}$ is the critical temperature for BEC.
\end{thebibliography}
\end{document}